\documentclass[sigconf, nonacm]{acmart}

\usepackage{pvldb}

\usepackage{algorithmic}
\usepackage{array}
\usepackage{multirow}
\usepackage{placeins}

\begin{document}

\title{MaDI-Bench: An End-to-End Data Integration Benchmark}

\author{Aaron Steiner}
\authornote{These authors contributed equally to this work.}
\orcid{0009-0006-6946-7057}
\affiliation{%
  \department{Data and Web Science Group}
  \institution{University of Mannheim}
  \city{Mannheim}
  \country{Germany}
}
\email{aaron.steiner@uni-mannheim.de}

\author{Ralph Peeters}
\authornotemark[1]
\orcid{0000-0003-3174-2616}
\affiliation{%
  \department{Data and Web Science Group}
  \institution{University of Mannheim}
  \city{Mannheim}
  \country{Germany}
}
\email{ralph.peeters@uni-mannheim.de}

\author{Christian Bizer}
\orcid{0000-0003-2367-0237}
\affiliation{%
  \department{Data and Web Science Group}
  \institution{University of Mannheim}
  \city{Mannheim}
  \country{Germany}
}
\email{christian.bizer@uni-mannheim.de}

\begin{abstract}
Data integration is the process of combining data from multiple, heterogeneous sources into a consistent, unified representation. Data integration involves a sequence of interdependent tasks including schema matching, value normalization, blocking, entity matching, and data fusion. Existing table-based benchmarks either evaluate these steps in isolation or cover only incomplete versions of the data integration pipeline, omitting specific steps. The lack of public end-to-end data integration benchmarks hinders research on data integration methods that address the integration process as a whole and account for the interdependencies among the different tasks. This paper fills this gap by introducing the Mannheim Data Integration Benchmark (MaDI-Bench), the first benchmark for the end-to-end integration of relational tables covering all steps of the integration process. MaDI-Bench contributes (i) a set of end-to-end data integration tasks spanning several application domains, each requiring the full schema matching, value normalization, entity matching, and data fusion pipeline, and (ii) a generic method for deriving task variants that mitigates rapid benchmark saturation as data integration systems advance. We validate the benchmark using human-engineered pipelines, a best-of-breed pipeline, an LLM workflow, and a pipeline written by a coding agent. The validation demonstrates the utility of the benchmark for measuring the step-wise as well as the end-to-end performance of data integration pipelines. All benchmark artifacts are available for public download.

\end{abstract}

\maketitle

\pagestyle{plain}

\section{Introduction}
\label{sec:intro}

Data integration~\cite{doan2012principles,dong2015big} combines data from multiple heterogeneous sources into a single, coherent representation. The interdependent steps include schema matching~\cite{bernstein2011schema}, value normalization, entity matching~\cite{christophides2020matchig}, and data fusion~\cite{bleiholder2009fusion}. These steps align source schemas and value formats, identify records describing the same real-world entity, and resolve conflicting values into one fused value per attribute. Each step has been studied in separation for decades~\cite{doan2012principles,dong2015big}, but in practice, these steps are tightly coupled: the output of one step constrains what the next step can achieve, and errors propagate through the pipeline. For example, successful value normalization makes the following entity matching task easier, unsuccessful entity matching results in multiple record clusters per real-world entity or overly large clusters describing multiple real-world entities. Both types of errors impact data fusion, as multiple clusters might end up as duplicate records in the output and as over-sized clusters covering different entities make conflict resolution in data fusion difficult. Therefore, evaluating the steps of the data integration process in isolation gives an incomplete picture of how a data integration system performs on the end-to-end data integration task that practitioners actually face.

Recent progress in large language models (LLMs) and agentic systems is changing what is feasible: LLM-based workflows can read heterogeneous inputs, propose schema mappings, normalize values, and select conflict resolution rules without human involvement~\cite{beyond-sql}. Agents can also collaborate with human data engineers using human-in the loop approaches~\cite{data-agents-survey,llm-data-survey,llm-ds-agent-survey}. This makes automated and human-AI \emph{end-to-end} data integration a realistic target. To drive and measure progress towards this goal, the community needs benchmarks that evaluate the whole integration pipeline, not just one step at a time.

Existing benchmarks do not provide for this need. Many benchmarks target a single step: schema matching~\cite{koutras2021valentine}, entity matching~\cite{mudgal2018deep,wdc-products}, and data fusion~\cite{yinTruthDiscoveryMultiple2007}. Each step is evaluated using different datasets and task-specific metrics. ETL and ELT benchmarks such as TPC-DI~\cite{tpc-di} and ELT-Bench~\cite{elt-bench} cover end-to-end transformation and loading workloads, but not end-to-end data integration tasks involving entity matching and conflict resolution.
What is missing is an end-to-end data integration benchmark that covers the steps of schema matching, value normalization, entity matching, and data fusion at different levels of difficulty across different application domains, and that supports the measurement of reference-free structural as well as ground-truth-based end-to-end metrics.

This paper presents the Mannheim Data Integration Benchmark \textbf{(MaDI-Bench)}, an end-to-end benchmark for evaluating data integration systems. MaDI-Bench provides integration tasks that take several heterogeneous source tables as input and require their integration into a single, consistent target table as output, exercising schema matching, value normalization, entity matching, and data fusion together. The MaDI tasks cover five application domains: Games, companies, music, products, and scientific papers. %
We also introduce a variant-generation method to provide additional easy, medium, and hard variants for each baseline task, so that integration methods can be tested for effectiveness and efficiency on tasks having different levels of difficulty.
Overall, MaDI-Bench offers 20 unique integration tasks across the five domains. To assess pipeline output, we organize evaluation metrics along three quality dimensions (coverage, consistency, and correctness) and three evaluation setups (reference-free structural, silver-standard, and ground-truth), including structural metrics such as record gain, density gain, output density, and fusion ratio. We validate MaDI-Bench using human-engineered PyDI\footnote{\url{https://github.com/wbsg-uni-mannheim/PyDI/tree/main}} pipelines, an LLM workflow~\cite{beyond-sql}, a best-of-breed pipeline that runs a committee of competing methods at each pipeline step and selects one winner per step, and a Claude Code agent that generates and executes integration code without relying on task-specific training data.

\noindent This paper makes the following contributions:

\begin{enumerate}
    \item A set of base end-to-end data integration tasks across five application domains (games, companies, music, products, and scientific papers), each spanning the subtasks schema matching, value normalization, entity matching, and data fusion. For each base task, we provide a gold schema mapping, training, validation, and test sets for entity matching, which also evaluate blocking, and validation and test sets for value normalization and data fusion. The test sets also serve as the ground truth for the end-to-end evaluation.
    \item A variant-generation method for deriving end-to-end data integration tasks at different levels of difficulty from the five base tasks. The variants of increased difficulty aim at making the benchmark more future-proof as data integration systems improve, while the easy variant supports the evaluation of simple but computationally efficient methods.
    \item We validate MaDI-Bench using four end-to-end data integration approaches: human-engineered pipelines, an LLM workflow, a best-of-breed pipeline that selects methods per integration step, and a Claude Code integration agent.
\end{enumerate}

\textbf{Resource Availability:} All artifacts together with the code for replicating the experiments are available in the MaDI-Bench repository\footnote{\url{https://github.com/wbsg-uni-mannheim/MaDI-Bench}}. Additional details and examples can be found on the benchmark's website\footnote{\url{https://wbsg-uni-mannheim.github.io/MaDI-Bench/}}.

\section{The Base Tasks}\label{sec:tasks}

\begin{figure*}[t]
  \centering
  \includegraphics[width=\textwidth]{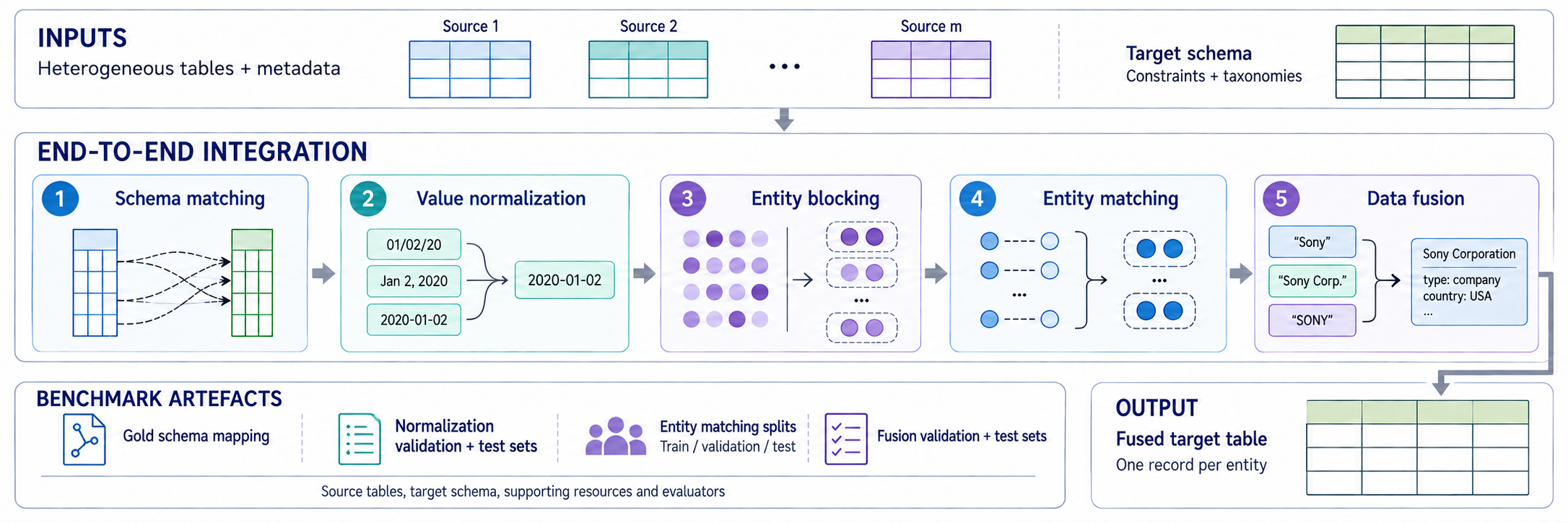}
  \caption{End-to-end data integration workflow and benchmark artefacts.}
  \Description{Source tables with metadata and a target schema feed a pipeline
  of schema matching, value normalization, entity blocking, entity matching,
  and data fusion.
  The pipeline produces a fused target table. Benchmark artifacts include
  schema mappings, normalization validation and test sets, labeled matching
  pairs, fusion validation and test sets,
  and supporting resources and evaluators.}
  \label{fig:workflow}
\end{figure*}

MaDI-Bench is built on five base end-to-end data integration tasks: Games, Companies, Music, Products, and Scientific Papers. Each of these base tasks requires an integration system to solve the full pipeline, from schema matching and value normalization through blocking and entity matching, to data fusion. As input, a system receives a set of source tables, metadata describing the tables, the target schema with its constraints and taxonomies, and optionally labeled training and validation sets. It must return a single fused table that conforms to the target schema and contains one record per real-world entity. For the step-wise evaluation, a system must additionally report its intermediate results: the proposed mapping from source attributes to the target schema, the normalized source tables, the candidate pairs of its blocker, and the matched record pairs. These are evaluated against the gold schema mapping, the normalization test set, and the entity matching test sets. Figure~\ref{fig:workflow} gives an overview over the inputs, the data integration tasks, and the benchmark artefacts. The artefacts and benchmark tasks are described in the following sections. Table~\ref{tab:task-stats} provides key statistics and an overview of the schemata of the datasets of the base tasks.

\subsection{Benchmark Artefacts}\label{sec:tasks-artifacts}

\begin{table*}[!t]
\centering
\caption{Statistics and attributes of the MaDI base task datasets. Identifier columns are excluded from attribute and density counts. Long attribute lists are truncated with \texttt{[...]}.}
\label{tab:task-stats}
\scriptsize
\setlength{\tabcolsep}{1.4pt}
\begin{tabular}{@{}llrrrrrr>{\raggedright\arraybackslash}p{9.8cm}@{}}
\toprule
Task & Source & Rows & Attrs. & Str & Num & Date & Density & Attributes \\
\midrule
\multirow{4}{*}{Games}
  & DBpedia & 46,580 & 6 & 5 & 0 & 1 & 90.9\% & title, launch\_yr, studio, system, genre, franchise \\
  & Metacritic & 20,494 & 8 & 5 & 2 & 1 & 97.7\% & game\_title, year\_published, made\_by, console, genres, press\_rating, player\_rating, age\_rating \\
  & Sales & 7,877 & 10 & 6 & 3 & 1 & 98.7\% & prod\_title, launch\_dt, studio, dist, hw, genre, press\_score, comm\_rating, age\_classification, units\_sold\_mm \\
\cmidrule(lr){2-9}
  & \emph{Target schema} & -- & 10 & 7 & 2 & 1 & -- & name, releaseYear, developer, genres, publisher, platform, criticScore, userScore, ESRB, series \\
\midrule
\multirow{4}{*}{Companies}
  & DBpedia & 10,085 & 8 & 5 & 2 & 1 & 62.2\% & org\_name, established, nation, headquarters, sector, keypeople\_name, total\_assets\_val, annual\_income \\
  & Forbes & 2,000 & 6 & 4 & 2 & 0 & 99.1\% & company, url, region, business\_segment, asset\_value, sales\_figure \\
  & FullContact & 1,931 & 5 & 4 & 0 & 1 & 61.9\% & Attribute\_2, Attribute\_3, Attribute\_4, Attribute\_5, Attribute\_6 \\
\cmidrule(lr){2-9}
  & \emph{Target schema} & -- & 8 & 5 & 2 & 1 & -- & name, founded, country, city, industry, assets, revenue, keypeople \\
\midrule
\multirow{4}{*}{Music}
  & Discogs & 22,627 & 8 & 6 & 1 & 1 & 93.1\% & title\_str, performer, pub\_dt, origin\_loc, duration, imprint, category, tracks\_track-name \\
  & Last.fm & 9,865 & 4 & 3 & 1 & 0 & 73.2\% & album\_title, band, album\_length, tracks\_track-name \\
  & MusicBrainz & 4,763 & 6 & 4 & 1 & 1 & 94.6\% & Attribute\_2, Attribute\_3, Attribute\_4, Attribute\_5, Attribute\_6, Attribute\_9 \\
\cmidrule(lr){2-9}
  & \emph{Target schema} & -- & 8 & 6 & 1 & 1 & -- & name, artist, release-date, release-country, label, genre, tracks, duration \\
\midrule
\multirow{5}{*}{Products}
  & Dataset 1 & 812 & 24 & 15 & 9 & 0 & 53.2\% & manufacturer, product\_name, product\_description, list\_price, currency\_code, product\_url, model\_name, manufacturer\_part\_number, category, gpu\_chipset, video\_memory\_gb, \texttt{[...]}\\
  & Dataset 2 & 812 & 24 & 15 & 9 & 0 & 52.5\% & brandName, name, descriptionText, priceAmount, currency, productUrl, modelName, mpn, productCategory, chipset, vramGb, capacityGb, readSpeedMbps, writeSpeedMbps, busType, interfaceType, \texttt{[...]} \\
  & Dataset 3 & 762 & 24 & 15 & 9 & 0 & 51.9\% & Brand, ProductTitle, Details, Price, Currency, Link, Model, PartNo, Type, Chipset, MemorySizeGB, \texttt{[...]} \\
  & Dataset 4 & 626 & 24 & 15 & 9 & 0 & 52.4\% & mfr, name, desc, amt, cur, link, mdl, pn, cat, chip, vram, cap\_gb, rd\_mbs, wr\_mbs, bus, iface, \texttt{[...]} \\
\cmidrule(lr){2-9}
  & \emph{Target schema} & -- & 24 & 15 & 9 & 0 & -- & brand, title, description, price, priceCurrency, url, model, model\_number, product\_type, chipset\_name, vram\_gb, storage\_gb, read\_speed\_mb\_s, write\_speed\_mb\_s, bus\_type, interface\_type, \texttt{[...]} \\
\midrule
\multirow{4}{*}{Papers}
  & Crossref & 60,749 & 13 & 10 & 2 & 1 & 85.9\% & work\_type, title\_text, contributor\_names, issued\_year, container\_title, publisher\_name, abstract\_text, volume\_id, issue\_id, page\_first, page\_last, reference\_total, cited\_total \\
  & DBLP & 60,591 & 9 & 8 & 0 & 1 & 68.1\% & entry\_type, publication\_title, author\_list, pub\_year, venue\_name, volume\_no, issue\_no, page\_start, page\_finish \\
  & OpenAlex & 60,719 & 12 & 9 & 2 & 1 & 88.6\% & work\_kind, display\_title, authors\_list, year\_published, source\_name, topic\_terms, volume\_tag, issue\_tag, start\_page, end\_page, refs\_count, citations\_count \\
\cmidrule(lr){2-9}
  & \emph{Target schema} & -- & 11 & 8 & 2 & 1 & -- & type, title, authors, publication\_year, journal, volume, issue, first\_page, last\_page, referenced\_works\_count, cited\_by\_count \\
\bottomrule
\end{tabular}
\end{table*}

Each benchmark task provides the source data, descriptions of that data, and the resources to evaluate every step of the pipeline. Across the five base tasks, these artefacts amount to more than 93{,}000 labeled record pairs for entity matching, 1{,}745 labeled normalization cells, 1{,}000 human-verified fusion records carrying close to 8{,}000 verified attribute values, and a gold schema mapping per task.

\textbf{Source tables and metadata:} Each task provides one table per source. Every table is accompanied by a metadata file that records provenance information of the data, the columns it contains, and the date it was published. The source publication dates span more than a decade, from the 2014 Forbes Global 2000 list to 2026 snapshots of DBLP, Crossref, and OpenAlex. We deliberately include dated sources, since outdated values are a recurring challenge in real integration settings. An advanced system can use the metadata during integration, for example, to weigh a source's trustworthiness from its provenance and publication date.

\textbf{Target schema:} Defines the columns that the output table (Table~\ref{tab:task-stats}) should contain. In addition to the column names, each attribute is given a value constraint that bounds its values, for example a company founding date between 1700-01-01 and 2016-12-31. Six categorical attributes in Games, Companies, and Music are tied to a taxonomy that ships with the task, such as the Global Industry Classification Standard\footnote{https://www.msci.com/indexes/index-resources/gics} for company industries, and the schema fixes the taxonomy level at which their values should be fused. The other categorical attributes are restricted to a list of allowed values, such as the product and publication types, or constrained by a consistency rule, such as expected value families for product bus types or ISO 4217 codes for currencies.

\textbf{Gold schema mapping:} Each task includes a gold mapping from the source attributes to the target schema, against which the correspondences proposed by a system are scored during the evaluation of the schema matching step.

\textbf{Normalization sets:} To evaluate value normalization, each task provides a normalization validation and test set. Each cell in these sets pairs a source value with its correctly normalized form. The cells are selected to represent the variety of the normalization challenges that appear in a task rather than their frequency: each distinct transformation enters a set at most five times. Most cells require a formatting rule or a mapping onto a vocabulary that the task declares, such as padding a partial date, converting between units of measurement, resolving an alias, or mapping a label onto one of the provided taxonomies. Some cells already carry the correct value and test whether a normalizer leaves correct values intact. A small share of the transformations requires external background knowledge, such as expanding a journal abbreviation. Table~\ref{tab:task-eval-resources} reports the number of cells in the validation and test sets.

\textbf{Entity matching sets:} Each task provides labeled training, validation, and test sets for entity matching. Table~\ref{tab:task-eval-resources} reports the number of correspondences (record pairs) between sources (e.g., MusicBrainz-Discogs and MusicBrainz-Last.fm for the Music task) and the share of positive pairs (matches) in the test set, which is between 25\% and 33\% across tasks. The sets concentrate on difficult to match corner cases near the decision boundary between matches and non-matches: true matches with dissimilar surface forms, and non-matches with high surface similarity~\cite{wdc-products}. To reduce selection bias, candidate pairs are drawn from several blocking strategies per task, including Jaccard-similarity and key-based blocking. The sets are disjoint at the pair level: no labeled pair appears in more than one set. We verified each test set by hand, except for Papers, whose matches are automatically derived from available DOI identifiers, and Products, which reuses the mapping of the WDC Products~\cite{wdc-products} benchmark from which the datasets were derived. The entity matching sets also evaluate blocking: training and validation sets can tune blockers for pair completeness and reduction ratio, and the test set provides the final evaluation.

\begin{table}[!h]
\centering
\caption{Labeled evaluation resources per base task.}
\label{tab:task-eval-resources}
\scriptsize
\setlength{\tabcolsep}{1.6pt}
\begin{tabular}{@{}lrrrrrrrrr@{}}
\toprule
 & \multicolumn{2}{c}{Normalization} & \multicolumn{5}{c}{Entity matching} & \multicolumn{2}{c@{}}{Fusion} \\
\cmidrule(lr){2-3}\cmidrule(lr){4-8}\cmidrule(l){9-10}
Task & Val. & Test & Src. pairs & Train & Val. & Test & Test +\% & Val. & Test \\
\midrule
Games     & 202 & 182 & 2 & 934    & 234    & 739    & 30.0\% & 775   & 793 \\
Companies & 91  & 92  & 2 & 1,971  & 948    & 599    & 29.4\% & 614   & 610 \\
Music     & 84  & 80  & 2 & 36,658 & 17,010 & 2,000  & 33.3\% & 727   & 726 \\
Products  & 235 & 250 & 3 & 4,484  & 599    & 599    & 25.0\% & 819   & 836 \\
Papers    & 281 & 248 & 2 & 10,666 & 2,666  & 13,332 & 25.0\% & 902   & 922 \\
\bottomrule
\end{tabular}
\end{table}

\textbf{Fusion sets:} The fusion evaluation sets are created by selecting entities whose records disagree on the values of the entity, ensuring that  conflict resolution has to take place. Human verifiers determined the correct value of each attribute through web research across trustworthy sources that are not part of the task sources. Every gold value is grounded in the actual real value for that entity instead of chosen from the values in the data sources. Disagreements between verifiers were resolved by gathering further evidence through a second round of web research. Each task in MaDI-Bench provides 100 validation and 100 test entities for data fusion. Table~\ref{tab:task-eval-resources} reports the number of target values they carry, excluding identifiers. The validation set allows for integration systems to tune their conflict-resolution strategy, while the test set is used to calculate the final fusion accuracy. The attribute-specific comparison functions that score fusion accuracy, for example a tolerance of $\pm$1 point for game critic scores or of 2\% for company revenue, are part of the evaluation package of the MaDI-Bench repository, which also scores the normalization test sets. Other implementations can be used, provided they follow the per-domain rule tables documented there. The following section introduces each task and its specific challenges.

\subsection{Benchmark Tasks}\label{sec:benchmark tasks}

Table~\ref{tab:task-stats} summarizes the source-level characteristics of the five base tasks. For each source, we report the number of rows, the number of non-identifier attributes, and the density, measured as the fraction of non-empty cells over these attributes. The target schema row reports the number of non-identifier attributes in the integrated output schema. The attributes column lists these attributes. %

\textbf{Games:} The Games task requires the integration of datasets from three video-game sources: Metacritic, a dataset about game sales, and games data from DBpedia. Metacritic focuses on review scores and ESRB ratings, the sales dataset contains commercial performance information, and DBpedia contributes encyclopedic attributes such as release dates, developers, platforms, genres, and series information. The target schema contains ten non-identifier attributes. The main challenge is that the same game on a different platform is a separate entity, so a matcher must not over-rely on title: identical titles recur across platforms, and sequels, special editions, and downloadable content differ only slightly in name. The task also requires the normalization of platform names and genres into the platform and genre taxonomies.

\textbf{Companies:} 
The Companies task requires the integration of records from the Forbes Global 2000 list, a DBpedia company extract, and a FullContact company-profile sample. Forbes provides financial figures, industries, and countries, DBpedia contributes founding dates, headquarters, industries, key people, and financial figures, and FullContact adds headquarters locations, founding dates, and key people. The target schema contains eight non-identifier attributes. The included companies span the globe, making normalization of legal suffixes and resolving their locations non-trivial. The difficulties of the task further include company-name variants, spelling differences, corporate hierarchies, and the normalization of financial figures, countries, and industry categories.

\textbf{Music:} 
The Music task integrates release-level records from Discogs, Last.fm, and MusicBrainz. Discogs provides release metadata with music labels, genres, countries, and track lists. Last.fm contributes album metadata, and MusicBrainz provides release titles, artists, dates, countries, durations, and track lists. The 37,255 input records describe albums, EPs, and singles, with partial overlap across sources. The target schema contains eight non-identifier attributes. The main challenge lies in heterogeneous value formats: for example, Discogs mostly records only the release year where MusicBrainz records the full date, and Discogs writes country abbreviations where MusicBrainz writes the full country name. Further difficulties involve title and artist variants, sparse source records, and normalization of dates, countries, and track lists.

\textbf{Products:} 
The Products task is based on a data sample of the entity matching benchmark WDC Products~\cite{wdc-products}. The product attribute values were extracted from free-text product offers using information extraction techniques. Product datasets cover GPUs, SSDs, HDDs, and USB sticks. The sources describe  product offers using different source-schema variants. The target schema contains 24 non-identifier attributes, making it the largest schema in the benchmark. The large schema makes matching delicate: a small difference in a single technical attribute can differentiate a match from a non-match, and the same attribute appears under different names using different value formats across sources. The task contains taxonomies for product types, storage interfaces, and GPU memory types, and the target schema fixes the unit of each capacity, speed, dimension, and weight attribute, for example gigabytes, megabytes per second, millimetres, and grams.

\textbf{Scientific Papers:} 
The Scientific Papers task integrates a selection of computer-science paper records from DBLP, Crossref, and OpenAlex. Each source annotates papers with bibliographic attributes such as title, authors, publication year, venue, page range, and citation-related counts. The target schema contains 11 non-identifier attributes. As the largest task by row count, Scientific Papers makes blocking efficiency a challenge: a blocker with a poor reduction ratio leads to hundreds of thousands of record pair comparisons. As a result, this task is especially suited for evaluating the efficiency of integration systems in addition to their effectiveness. Challenges include title and author-list variants, incomplete identifiers, publication-type and venue normalization, and sparse metadata for volume, issue, pages, and citation counts. 

\section{Experimental Validation}\label{sec:eval}

This section validates MaDI-Bench using four end-to-end data integration pipelines. Section~\ref{sec:eval-pipelines} describes the different pipelines, Section~\ref{sec:metrics} introduces the metrics that we use for the evaluation, and Section~\ref{sec:eval-base} reports per-step as well as end-to-end results.

\subsection{End-to-End Integration Pipelines}\label{sec:eval-pipelines}

We validate the benchmark using four end-to-end data integration pipelines. The pipelines span the design space from completely hand-engineered to fully LLM-driven. The human-designed pipeline (P1) and the LLM workflow (P3) build on the PyDI data integration framework\footnote{\url{https://github.com/wbsg-uni-mannheim/PyDI}}. Both have access to the same integration functions and differ in their configuration and supervision protocols. The best-of-breed pipeline (P2) chooses, for each task step, between specialized methods from the literature and methods of PyDI. The coding agent Claude Code (P4) generates and executes task-specific integration code without PyDI\@.

\textbf{(P1) Human-designed PyDI pipeline:} P1 consists of specific workflows for each of the five base tasks. The workflows are all implemented using the PyDI data integration framework and were manually engineered by master and PhD students of the University of Mannheim. 
The workflows were reviewed and improved in several iterations.
Each workflow follows the same order: The data engineers run the PyDI LLM schema matcher, inspect the generated correspondences, and hand-correct missed or misplaced attributes before translating the sources into the target schema. Afterwards, a normalization step aligns value formats, including date and number parsing, unit conversion, and selected taxonomy mappings. In Games, for example, this includes mapping platform variants to the target taxonomy. Blocking uses either embedding-based blockers or task-specific blocking keys. The validation target for blocking is at least 97\% pair completeness, meaning that at least 97\% of labeled matches remain in the candidate set. Among blockers that reach this floor, engineers prefer the one with the highest reduction ratio.

For entity matching, engineers compare rule-based matchers and train tree-based matchers using similarity features. They further inspect validation errors, and apply global one-to-one matching algorithms to the results if deemed necessary. For fusion, they choose PyDI conflict-resolution heuristics per attribute based on data type, domain knowledge, and validation set results. The employed heuristics include voting, source trust, longest or shortest string, numeric aggregation, and set union for list-valued attributes.
P1 is a practical human reference point which does not claim optimality. Its fused output also serves as the silver reference (Section~\ref{sec:metrics}).

\textbf{(P2) Best-of-breed (BoB) pipeline:} The BoB pipeline chains strong specialized components per integration step to achieve a good end-to-end integration result. It executes the five integration steps without backwards iteration. At each step, a committee of competing methods is executed, the result of every method is scored using the validation set, and the winner's output is passed to the next step. The final pipeline is the chained composition of the per-step winners and emits a single output table.

P2 uses the committees of Table~\ref{tab:committees} (Section~\ref{sec:taskgen:validation}). The schema matching committee consists of COMA~\cite{coma}, which combines several label- and instance-based sub-matchers, the Magneto~\cite{magneto} matcher that pairs a small retrieval model with an LLM reranker, a Sentence-BERT embedding matcher~\cite{sbert} based on nearest-neighbor retrieval of column embeddings, and the label-based, instance-based, duplicate-based, and LLM matchers of PyDI.

The normalization committee consists of two members. The first leaves the source values unchanged. The second applies, for each attribute, one of several type-specific rules: text cleaning for strings, date parsing for dates and years, locale-aware number parsing, and country-code or taxonomy lookups for categorical values.

The blocking committee consists of standard blocking, token blocking and sorted-neighborhood blocking~\cite{christophides2020matchig}, BM25 blocking~\cite{bm25}, Sentence-BERT embedding-based~\cite{sbert} blocking, and SC-Block~\cite{scblock}, a supervised contrastive blocker. The committee keeps, for each pair of source datasets, the blocker with the highest reduction ratio among those that reach a pair completeness of 97\% on the entity matching validation set. Blockers that produce more than five million candidate pairs are excluded. If no blocker reaches the floor, the blocker with the highest pair completeness is selected.

The entity matching committee consists of Ditto~\cite{ditto}, a PLM-based matcher, and Magellan~\cite{magellan}, a matcher that automatically creates and selects similarity-based features for a random forest matcher. 
The fusion committee consists of PyDI conflict-resolution heuristics and learned truth-discovery models. Each heuristic resolves a conflict by a fixed rule such as majority voting, source trust, longest string, median, or set union. A per-attribute selector chooses for each attribute the heuristic that scores best on the fusion validation set~\cite{bleiholder2009fusion}. Majority voting and source trust are additional uniform baselines that apply a single rule to every attribute. The other members are truth-discovery models that estimate source trust: TruthFinder~\cite{yinTruthDiscoveryMultiple2007}, LTM~\cite{zhaoBayesianApproachDiscovering2012}, AccuSim~\cite{dongIntegratingConflictingData2009}, CASEFusion~\cite{lyuTruthDiscoveryClaim2017}, and FusionQuery~\cite{zhuFusionQueryOndemandFusion2024}.

\textbf{(P3) LLM workflow:} P3 is the end-to-end data integration workflow described in~\cite{beyond-sql}. It uses the same methods from the PyDI framework as the human pipeline, but an LLM configures them.

The LLM performs schema matching by calling the LLM schema matcher. For the other steps it configures non-LLM methods instead of integrating the records itself, allowing the workflow to scale. The workflow receives the source tables, a target schema, the benchmark taxonomies, and the fusion validation set of the benchmark. It matches the schemas using GPT-5.5 and normalizes the values by mapping them to the predefined taxonomies. Non-categorical attributes are normalized by a column profiler that selects a semantically correct normalization function. An embedding-based blocker then reduces the sources to candidate pairs. From a sample of these pairs, the workflow builds its own machine-labeled entity matching training and validation sets~\cite{beyond-sql}, using GPT-5.2 as the labeler. These machine-labeled sets are independent of the sets that are provided by this benchmark. Using these artefacts, the LLM trains an entity matcher, choosing among rule-based matchers and feature-based models such as random forest and XGBoost. For data fusion, the model determines a configuration which assigns each attribute a conflict-resolution heuristic from the set of heuristics implemented in PyDI, for example longest string, voting, or source-based trust. The workflow builds several such configurations, some of them tuned on the fusion validation set, and keeps the one with the highest accuracy on that set.

\textbf{(P4) Claude Code (CC) pipeline:} P4 uses the coding agent Claude Code 2.1.280 with Claude Opus 5.5 to write and execute its own integration pipeline for each task. The agent runs fully unsupervised in an isolated workspace. It receives the source tables with their metadata files, the target schema, the benchmark taxonomies, and a short description of the definition of an entity for each domain. Each run is limited to 18 hours of wall-clock time and 100 US dollars of LLM spend. PyDI and its documentation are unavailable, and web search and web fetching are disabled. An OpenAI embeddings endpoint is available; the agent used it only on the three Companies variants. Instead, the agent receives seven free text skill files for schema matching, normalization, blocking, entity matching, clustering, fusion, and pipeline diagnosis. Each skill describes possible methods, typical errors, and checks for its step. The authors wrote the skills, which are available in our repository\footnote{\url{https://github.com/wbsg-uni-mannheim/MaDI-Bench/tree/main/baselines/claude-code/skills}}.

The agent reads the relevant skills and implements the complete integration process itself from scratch. It inspects the sources, maps their attributes to the target schema, and writes code for normalization, blocking, matching, and fusion. It then executes the code, inspects errors and diagnostics such as missing values, schema conformity, and cluster sizes, and iteratively revises the code until it considers the task solved or reaches its allowed budget (which did not happen in our testing). P4 receives no task specific training and validation data.

\subsection{Evaluation Metrics}
\label{sec:metrics}

The MaDI-Bench evaluation suite calculates both step-level and end-to-end evaluation metrics. For the individual pipeline steps, we use the established metrics: Schema matching is evaluated as the F1 score over predicted attribute correspondences. Value normalization is evaluated as the accuracy over the cells of the normalization test set. Blocking is evaluated with \emph{pair completeness} and \emph{reduction ratio}.
For entity matching we report F1 on the test set. Data fusion uses \emph{accuracy}: the share of the ground-truth cells for which the fused output holds a correct value, pooled over all attributes. A value is correct when the attribute's comparator accepts it, for example within a numeric tolerance or as an equal set for list attributes. A missing value or entity counts as wrong.

Beyond the step-wise metrics, we calculate end-to-end metrics across the full data integration process, organized along two orthogonal axes: the \emph{quality dimension} and the \emph{reference level}.

The quality dimension captures three notions of output quality: \emph{Coverage} measures whether the fused output contains the relevant amounts of entities and values. It combines four reference-free signals computed on the fused output and the input sources with three metrics that compare the output's entities and values against a reference table. \emph{Consistency} measures whether the output conforms to the target schema and the expected value formats. Both consistency metrics build on a per-cell schema validity check, reported once on the output alone and once as a difference against a reference. \emph{Correctness} measures whether the matched records and fused values are actually correct and is therefore defined only against a reference. Its metrics separate cluster-level from value-level agreement: output and reference clusters are aligned by best record overlap, and the value metrics are computed on aligned pairs.

The second axis reflects the reliance on \emph{reference data} for the evaluation. We distinguish three levels of reference: \emph{Reference-free} metrics characterize the output of a pipeline on its own. They require no labels and are always available, but can only describe the structure of the output, not its correctness. \emph{Ground-truth-reference} metrics compare against the test sets provided together with the benchmark, which provide the highest level of correctness but are calculated only on the small slice of the output that is covered by the test sets. \emph{Silver-reference} metrics compare the fused output against the output values of a reference pipeline. In this paper, we use the output of the human-engineered PyDI pipeline (P1, Section~\ref{sec:eval}) as silver standard. Using the output of a reference pipeline as silver standard  results in broader evaluation coverage than the ground-truth reference, but offers weaker correctness guarantees. Table~\ref{tab:e2e-metrics} defines the metrics and the reference levels at which they apply.

\begin{table}[t]
\centering
\caption{End-to-end metrics by quality dimension. Ref.\ lists the reference levels at which a metric applies: reference-free (RF), ground truth (GT), and silver (S).}
\label{tab:e2e-metrics}
\scriptsize
\renewcommand{\arraystretch}{1.15}
\setlength{\tabcolsep}{3pt}
\begin{tabular}{@{}llp{5.0cm}@{}}
\toprule
Metric & Ref. & Definition \\
\midrule
\multicolumn{3}{@{}l}{\textbf{Coverage}} \\
Record Gain & RF & Relative change in the number of output records compared to the largest input source. \\
Density Gain & RF & Non-null cell density of the fused table compared to the largest input source (by rows). \\
Output Density & RF & Fraction of non-null cells in the output table. \\
Fusion Ratio & RF & Share of output records assembled from two or more source records, i.e., how much cross-source fusion occurred. \\
Entity recovery & GT, S & Fraction of reference entities recovered as a single output cluster with exact overlap (Jaccard $=1$). \\
Value drift & GT, S & Mean per-attribute distance between output and reference value distributions: Jensen-Shannon~\cite{lin1991divergence} for categorical and text, normalized Wasserstein~\cite{ramdas2017wasserstein} for numeric and date attributes. \\
Value-density $\Delta$ & GT, S & Mean per-attribute difference between the output's and the reference's share of filled values. \\
\midrule
\multicolumn{3}{@{}l}{\textbf{Consistency}} \\
Schema validity & RF & Fraction of filled cells satisfying the target schema's constraints on their attribute: data type, numeric range, enumeration, string pattern or format, and taxonomy membership. \\
Schema validity $\Delta$ & GT, S & Output's schema-validity rate minus the reference's. Positive values mean the output conforms more closely than the reference. \\
\midrule
\multicolumn{3}{@{}l}{\textbf{Correctness}} \\
BCubed precision & GT, S & Cluster purity: per-record fraction of an output cluster's members that share the record's reference cluster, averaged over records. Low values show over-merging. \\
BCubed recall & GT, S & Cluster completeness: per-record fraction of a reference cluster's members that land in the record's output cluster, averaged over records. Low values show over-splitting. \\
BCubed F1 & GT, S & Harmonic mean of BCubed precision and recall, the main measure of clustering agreement. \\
Fusion accuracy & GT, S & Share of the reference cells whose fused value is correct, pooled over all attributes. A fused cell is correct when it matches the aligned reference value under the attribute's evaluation function. \\
Fully-correct rate & GT, S & Fraction of aligned output clusters whose fused record matches the reference on every evaluated attribute simultaneously, the strictest measure. \\
\bottomrule
\end{tabular}
\end{table}

\subsection{Validation Results for the Base Tasks}\label{sec:eval-base}

We run all pipelines on the five base tasks of MaDI-Bench. P1 and P2 are configured on the provided validation sets. When a component needs training data, they use the benchmark-provided training sets. P3 generates its own entity matching training and validation sets, following~\cite{beyond-sql}, and selects its fusion configuration on the fusion validation sets of the benchmark. P4 uses no task specific. Each pipeline is run once per task. P1 runs the PyDI LLM schema matcher with GPT-5. P2 uses GPT-5.4-mini for the LLM-based schema matchers. P3 uses GPT-5.5 for schema matching, taxonomy mapping, and the fusion configuration, and GPT-5.2 for labeling. P4 uses Claude Opus 5.5 for all steps.

\subsubsection{Schema Matching}\label{sec:eval-schema}

Table~\ref{tab:schema-results} reports schema matching F1 per task, with PyDI's label- and instance-based matchers as reference.

All four pipelines reach at least 92.68 F1 on every task, and 15 of the 20 pipeline scores are 100. The two reference matchers fail in complementary places. Label-based matching peaks at 73.8 on Papers, where the source headers stay descriptive, and falls to 26.1 on Companies, where FullContact uses anonymized columns such as \texttt{Attribute\_2} (Table~\ref{tab:task-stats}). Instance-based matching shows the reverse profile: 64.5 on Music, where it infers the equally opaque MusicBrainz columns from their values, but 7.7 on Products, possibly because several of its numeric attributes, such as read and write speeds, draw values from similar ranges. The Products headers such as \texttt{readSpeedMbps} and \texttt{rd\_mbs} stay descriptive enough for label-based matching to reach 61.6. The base tasks combine both naming and value heterogeneity, and neither reference matcher reaches more than 74 F1 on any task.

\begin{table}[t]
\centering
\caption{Schema matching F1 (\%) on the base tasks.}
\label{tab:schema-results}
\scriptsize
\setlength{\tabcolsep}{1.5pt}
\resizebox{\columnwidth}{!}{%
\begin{tabular}{@{}lcccccc@{}}
\toprule
Task & \shortstack{Human\\(P1)} & \shortstack{BoB\\(P2)} & \shortstack{LLM\\(P3)} & \shortstack{CC\\(P4)} & \shortstack{Label-based\\schema matching} & \shortstack{Instance-based\\schema matching} \\
\midrule
Games & 100.00 & 100.00 & 100.00 & 93.88 & 30.00 & 60.50 \\
Companies & 95.00 & 97.67 & 92.68 & 100.00 & 26.10 & 53.80 \\
Music & 100.00 & 100.00 & 100.00 & 100.00 & 28.60 & 64.50 \\
Products & 100.00 & 100.00 & 100.00 & 100.00 & 61.60 & 7.70 \\
Papers & 95.38 & 100.00 & 100.00 & 100.00 & 73.80 & 33.30 \\
\bottomrule
\end{tabular}}
\end{table}

\subsubsection{Value Normalization}\label{sec:eval-norm}

Table~\ref{tab:normalization-results} reports normalization accuracy on the normalization test sets (Table~\ref{tab:task-eval-resources}). The accuracy covers both cells that require a transformation and cells whose value is already correct, so a normalizer that changes correct values loses accuracy.

Normalization accuracy ranges from 6.04 to 97.83 across pipelines and tasks, and every pipeline scores below 40 on at least one task (Table~\ref{tab:normalization-results}). Averaged over the tasks, it is the lowest step score of every pipeline, between 28.45 (P2) and 60.12 (P4). Games is the hardest task for P1, P2, and P4, which stay below 9\%. Of its 182 test cells, 171 require a transformation, mostly a mapping of genres and platform aliases onto the target taxonomies. The best pipeline differs by task: P3 leads on Games and Music, and P4 on Companies, Products, and Papers. A possible explanation for the lead of P3 on Games is its explicit taxonomy-matching step, which maps source terms onto the provided taxonomies (Section~\ref{sec:eval-pipelines}).

\begin{table}[t]
\centering
\caption{Normalization accuracy (\%) on the normalization test sets of the base tasks.}
\label{tab:normalization-results}
\footnotesize
\setlength{\tabcolsep}{3.5pt}
\begin{tabular*}{\columnwidth}{@{\extracolsep{\fill}}lcccc@{}}
\toprule
Task & \shortstack{Human\\(P1)} & \shortstack{BoB\\(P2)} & \shortstack{LLM\\(P3)} & \shortstack{CC\\(P4)} \\
\midrule
Games & 6.59 & 6.04 & 37.91 & 8.79 \\
Companies & 34.78 & 29.35 & 84.78 & 97.83 \\
Music & 63.75 & 55.00 & 75.00 & 56.25 \\
Products & 64.00 & 22.00 & 30.00 & 73.20 \\
Papers & 52.02 & 29.84 & 29.84 & 64.52 \\
\bottomrule
\end{tabular*}
\end{table}

\subsubsection{Blocking and Entity Matching}\label{sec:eval-em}

Table~\ref{tab:blocking-em-results} reports blocking pair completeness (PC) and reduction ratio (RR) together with entity matching F1 per task, averaged over the source pairs. Pair completeness lies between 93.33 and 100 for all pipelines and tasks, close to or above the 97 floor that P1 and P2 tune for. %
Reduction ratios exceed 97\% on every task except Products, where P1 removes 71.89\% of the possible record pairs, which is feasible due to the size of the tasks sources, and the other pipelines between 96.50\% and 97.85\%. On Papers, the largest task, every pipeline removes at least 99.97\% of the possible record pairs, as a lower ratio would make the matching workload prohibitive. The resulting candidate set is also a main driver of the runtime (Section~\ref{sec:eval-e2e}).

Entity matching F1 ranges from 64.98 to 99.83 (Table~\ref{tab:blocking-em-results}) across tasks. On Companies, Music, and Papers, the four pipelines lie within 5.8 F1 per task. Games and Products show larger differences. On Games, P1 (89.47) and P4 (85.40) lead P2 and P3 by more than 18 F1. A possible explanation lies in the corner cases of the Games test set: identical titles recur across platforms, and the same game on another platform is a different entity (Section~\ref{sec:benchmark tasks}), which penalizes matchers that rely on title similarity. On Products, P4 (88.48) and P2 (85.48) lead P1 and P3 by 18 to 22 F1. P2 uses the PLM-based Ditto matcher. No pipeline achieves the best score on every task. The scores lie in the range that current matchers report on established benchmarks such as WDC Products~\cite{wdc-products}, DBLP-ACM, and DBLP-Scholar~\cite{ditto}. The lower matching F1 of P1 and P3 on Products resurfaces in the fusion results (Section~\ref{sec:eval-fusion}).

\begin{table}[t]
\centering
\caption{Blocking pair completeness (PC), reduction ratio (RR), and entity matching F1 (\%).}
\label{tab:blocking-em-results}
\scriptsize
\setlength{\tabcolsep}{2pt}
\resizebox{\columnwidth}{!}{%
\begin{tabular}{@{}lcccccccccccc@{}}
\toprule
& \multicolumn{4}{c}{PC (\%)} & \multicolumn{4}{c}{RR (\%)} & \multicolumn{4}{c@{}}{Matching F1 (\%)} \\
\cmidrule(lr){2-5}\cmidrule(lr){6-9}\cmidrule(l){10-13}
Task & P1 & P2 & P3 & P4 & P1 & P2 & P3 & P4 & P1 & P2 & P3 & P4 \\
\midrule
Games & 99.53 & 98.58 & 96.85 & 97.21 & 99.92 & 99.88 & 99.96 & 99.99 & 89.47 & 67.30 & 64.98 & 85.40 \\
Companies & 97.88 & 100.00 & 100.00 & 99.12 & 99.30 & 97.16 & 99.40 & 99.82 & 86.84 & 89.29 & 92.63 & 91.20 \\
Music & 96.55 & 99.85 & 99.85 & 100.00 & 99.67 & 99.64 & 99.85 & 99.62 & 95.37 & 94.84 & 98.10 & 95.48 \\
Products & 100.00 & 100.00 & 93.33 & 100.00 & 71.89 & 96.50 & 97.54 & 97.85 & 67.94 & 85.48 & 66.00 & 88.48 \\
Papers & 97.90 & 99.55 & 97.69 & 99.88 & 99.98 & 99.97 & 99.97 & 99.99 & 98.31 & 99.70 & 96.83 & 99.83 \\
\bottomrule
\end{tabular}}
\end{table}

\subsubsection{Data Fusion}\label{sec:eval-fusion}

Table~\ref{tab:fusion-results} reports fusion accuracy on the human-verified fusion test sets. A fused value can be wrong because matching kept the record with the correct value out of the entity's cluster, because matching added records of other entities, or because conflict resolution chose a wrong value. Fusion accuracy ranges from 45.02 to 91.11. P4 reaches the highest accuracy on Games, Companies, Music, and Products, 3.5 to 6.4 points above the second best pipeline. P1 leads on Papers with 91.11. The best score per task leaves between 8.89 (Papers) and 22.19 (Games) points open.

Products illustrates how matching errors propagate into fusion. The two pipelines with the highest Products matching F1, P4 and P2, also reach the highest Products fusion accuracy (82.78 and 79.31), since a missed correspondence or a wrong match bounds what fusion can recover. An example from the human workflow is the Kingston DataTraveler 100 G3 flash drive. It appears in all four sources, but the matcher links only the source~1 record with a read speed of 100~MB/s, while the verified value of 130 sits in an unmatched record. No conflict-resolution rule can recover a value that never enters the cluster. 

\begin{table}[t]
\centering
\caption{Data fusion accuracy (\%) on the base tasks.}
\label{tab:fusion-results}
\setlength{\tabcolsep}{2.5pt}
\begin{tabular}{@{}lcccc@{}}
\toprule
Task & \shortstack{Human\\(P1)} & \shortstack{BoB\\(P2)} & \shortstack{LLM\\(P3)} & \shortstack{CC\\(P4)} \\
\midrule
Games & 53.29 & 65.70 & 71.37 & 77.81 \\
Companies & 60.47 & 66.39 & 75.74 & 81.97 \\
Music & 45.02 & 81.40 & 75.21 & 86.64 \\
Products & 71.12 & 79.31 & 61.48 & 82.78 \\
Papers & 91.11 & 90.24 & 78.42 & 85.57 \\
\bottomrule
\end{tabular}
\end{table}

\subsubsection{End-to-End Results}\label{sec:eval-e2e}
\begin{table*}[t]
\centering
\caption{End-to-end metrics for the base tasks. Silver reference: the clusters with at least two source records in the P1 output. $\uparrow$/$\downarrow$ give the preferred direction.}
\label{tab:e2e-detailed}
\scriptsize
\renewcommand{\arraystretch}{1}
\setlength{\tabcolsep}{2pt}
\resizebox{\textwidth}{!}{%
\begin{tabular}{@{}lll *{6}{ccc}@{}}
\toprule
& Dimension & Metric
& \multicolumn{3}{c}{Games}
& \multicolumn{3}{c}{Companies}
& \multicolumn{3}{c}{Music}
& \multicolumn{3}{c}{Products}
& \multicolumn{3}{c}{Papers}
& \multicolumn{3}{c}{Mean} \\
\cmidrule(lr){4-6}\cmidrule(lr){7-9}\cmidrule(lr){10-12}\cmidrule(lr){13-15}\cmidrule(lr){16-18}\cmidrule(lr){19-21}
& & & \shortstack{BoB\\(P2)} & \shortstack{LLM\\(P3)} & \shortstack{CC\\(P4)} & \shortstack{BoB\\(P2)} & \shortstack{LLM\\(P3)} & \shortstack{CC\\(P4)} & \shortstack{BoB\\(P2)} & \shortstack{LLM\\(P3)} & \shortstack{CC\\(P4)} & \shortstack{BoB\\(P2)} & \shortstack{LLM\\(P3)} & \shortstack{CC\\(P4)} & \shortstack{BoB\\(P2)} & \shortstack{LLM\\(P3)} & \shortstack{CC\\(P4)} & \shortstack{BoB\\(P2)} & \shortstack{LLM\\(P3)} & \shortstack{CC\\(P4)} \\
\midrule
\multirow{5}{*}{\rotatebox[origin=c]{90}{\textbf{Ref-Free}}}
 & \multirow{4}{*}{Coverage} & Record Gain & $+0.40$ & $+0.32$ & $+0.09$ & $+0.19$ & $+0.28$ & $+0.24$ & $+0.38$ & $+0.37$ & $+0.31$ & $+0.17$ & $+0.39$ & $+0.08$ & $+0.24$ & $+0.37$ & $+0.25$ & $+0.28$ & $+0.35$ & $+0.19$ \\
 &  & Density Gain $\uparrow$ & $-0.25$ & $-0.26$ & $-0.24$ & $-0.05$ & $-0.02$ & $-0.02$ & $-0.18$ & $-0.14$ & $-0.17$ & $+0.06$ & $+0.04$ & $+0.06$ & $-0.03$ & $-0.05$ & $-0.03$ & $-0.09$ & $-0.09$ & $-0.08$ \\
 &  & Output Density $\uparrow$ & 0.66 & 0.65 & 0.67 & 0.57 & 0.60 & 0.61 & 0.81 & 0.84 & 0.81 & 0.57 & 0.55 & 0.57 & 0.88 & 0.86 & 0.88 & 0.70 & 0.70 & 0.71 \\
 &  & Fusion Ratio & 0.11 & 0.17 & 0.31 & 0.13 & 0.08 & 0.11 & 0.14 & 0.15 & 0.20 & 0.84 & 0.69 & 0.94 & 0.74 & 0.66 & 0.73 & 0.39 & 0.35 & 0.46 \\
\cmidrule(lr){2-21}
 & Consistency & Schema validity $\uparrow$ & 1.00 & 1.00 & 1.00 & 0.94 & 0.99 & 1.00 & 0.84 & 0.97 & 1.00 & 1.00 & 0.95 & 1.00 & 0.98 & 0.98 & 0.98 & 0.95 & 0.98 & 1.00 \\
\midrule
\multirow{9}{*}{\rotatebox[origin=c]{90}{\textbf{Silver Reference}}}
 & \multirow{3}{*}{Coverage} & Entity recovery $\uparrow$ & 0.50 & 0.41 & 0.78 & 0.37 & 0.33 & 0.31 & 0.76 & 0.77 & 0.78 & 0.22 & 0.05 & 0.26 & 0.93 & 0.82 & 0.92 & 0.56 & 0.47 & 0.61 \\
 &  & Value drift $\downarrow$ & 0.20 & 0.26 & 0.22 & 0.35 & 0.41 & 0.41 & 0.39 & 0.47 & 0.47 & 0.07 & 0.13 & 0.09 & 0.01 & 0.00 & 0.01 & 0.20 & 0.25 & 0.24 \\
 &  & Value-density $\Delta$ & $-0.23$ & $-0.23$ & $-0.22$ & $-0.30$ & $-0.27$ & $-0.26$ & $-0.11$ & $-0.07$ & $-0.10$ & $-0.02$ & $-0.04$ & $-0.02$ & $-0.01$ & $-0.04$ & $-0.02$ & $-0.13$ & $-0.13$ & $-0.12$ \\
\cmidrule(lr){2-21}
 & Consistency & Schema validity $\Delta$ $\uparrow$ & $+0.00$ & $+0.00$ & $+0.00$ & $+0.08$ & $+0.13$ & $+0.13$ & $-0.10$ & $+0.03$ & $+0.06$ & $+0.00$ & $-0.05$ & $+0.00$ & $-0.00$ & $-0.00$ & $+0.00$ & $-0.01$ & $+0.02$ & $+0.04$ \\
\cmidrule(lr){2-21}
 & \multirow{5}{*}{Correctness} & BCubed precision $\uparrow$ & 1.00 & 0.94 & 1.00 & 0.95 & 0.99 & 0.99 & 1.00 & 1.00 & 1.00 & 0.68 & 0.51 & 0.68 & 0.99 & 0.99 & 0.99 & 0.92 & 0.89 & 0.93 \\
 &  & BCubed recall $\uparrow$ & 0.63 & 0.57 & 0.81 & 0.66 & 0.63 & 0.62 & 0.69 & 0.70 & 0.71 & 0.60 & 0.43 & 0.64 & 0.97 & 0.91 & 0.97 & 0.71 & 0.65 & 0.75 \\
 &  & BCubed F1 $\uparrow$ & 0.77 & 0.71 & 0.90 & 0.77 & 0.77 & 0.76 & 0.82 & 0.82 & 0.83 & 0.64 & 0.47 & 0.66 & 0.98 & 0.95 & 0.98 & 0.80 & 0.74 & 0.82 \\
 &  & Fusion acc.\ $\uparrow$ & 0.79 & 0.68 & 0.90 & 0.55 & 0.51 & 0.48 & 0.71 & 0.71 & 0.62 & 0.58 & 0.45 & 0.57 & 0.91 & 0.74 & 0.85 & 0.71 & 0.62 & 0.69 \\
 &  & Fully-correct rate $\uparrow$ & 0.29 & 0.08 & 0.46 & 0.11 & 0.02 & 0.03 & 0.13 & 0.13 & 0.05 & 0.03 & 0.00 & 0.02 & 0.55 & 0.01 & 0.13 & 0.22 & 0.05 & 0.14 \\
\midrule
\multirow{9}{*}{\rotatebox[origin=c]{90}{\shortstack{\textbf{Ground Truth}\\\textbf{Reference}}}}
 & \multirow{3}{*}{Coverage} & Entity recovery $\uparrow$ & 0.46 & 0.31 & 0.85 & 0.91 & 0.77 & 0.81 & 0.87 & 0.91 & 0.96 & 0.59 & 0.08 & 0.76 & 1.00 & 0.83 & 0.99 & 0.77 & 0.58 & 0.87 \\
 &  & Value drift $\downarrow$ & 0.40 & 0.34 & 0.41 & 0.63 & 0.54 & 0.45 & 0.61 & 0.55 & 0.51 & 0.28 & 0.33 & 0.25 & 0.38 & 0.37 & 0.37 & 0.46 & 0.43 & 0.40 \\
 &  & Value-density $\Delta$ & $-0.19$ & $-0.20$ & $-0.18$ & $-0.18$ & $-0.15$ & $-0.15$ & $-0.10$ & $-0.07$ & $-0.10$ & $+0.04$ & $+0.02$ & $+0.05$ & $-0.03$ & $+0.03$ & $+0.04$ & $-0.09$ & $-0.08$ & $-0.07$ \\
\cmidrule(lr){2-21}
 & Consistency & Schema validity $\Delta$ $\uparrow$ & $+0.13$ & $+0.13$ & $+0.13$ & $-0.03$ & $+0.02$ & $+0.03$ & $-0.16$ & $-0.03$ & $+0.00$ & $-0.00$ & $-0.05$ & $-0.00$ & $-0.02$ & $-0.02$ & $-0.02$ & $-0.02$ & $+0.01$ & $+0.03$ \\
\cmidrule(lr){2-21}
 & \multirow{5}{*}{Correctness} & BCubed precision $\uparrow$ & 1.00 & 1.00 & 1.00 & 1.00 & 1.00 & 1.00 & 1.00 & 1.00 & 1.00 & 0.98 & 0.90 & 0.99 & 1.00 & 1.00 & 1.00 & 1.00 & 0.98 & 1.00 \\
 &  & BCubed recall $\uparrow$ & 0.67 & 0.56 & 0.89 & 0.95 & 0.88 & 0.90 & 0.93 & 0.96 & 0.98 & 0.81 & 0.44 & 0.90 & 1.00 & 0.92 & 1.00 & 0.87 & 0.75 & 0.93 \\
 &  & BCubed F1 $\uparrow$ & 0.80 & 0.71 & 0.94 & 0.97 & 0.93 & 0.95 & 0.97 & 0.98 & 0.99 & 0.89 & 0.59 & 0.94 & 1.00 & 0.96 & 1.00 & 0.92 & 0.83 & 0.96 \\
 &  & Fusion acc.\ $\uparrow$ & 0.66 & 0.71 & 0.78 & 0.66 & 0.76 & 0.82 & 0.81 & 0.75 & 0.87 & 0.79 & 0.61 & 0.83 & 0.90 & 0.78 & 0.86 & 0.77 & 0.72 & 0.83 \\
 &  & Fully-correct rate $\uparrow$ & 0.00 & 0.10 & 0.08 & 0.02 & 0.21 & 0.32 & 0.21 & 0.04 & 0.36 & 0.26 & 0.10 & 0.31 & 0.36 & 0.07 & 0.16 & 0.17 & 0.10 & 0.25 \\
\bottomrule
\end{tabular}}
\end{table*}

Table~\ref{tab:e2e-detailed} reports the end-to-end metrics for P2, P3, and P4 in the reference-free, silver-reference, and ground-truth views. P1 provides the silver reference and is not reported.

\textbf{Reference-free:} All three pipelines expand the largest input source, with mean record gains of 0.28 for P2, 0.35 for P3, and 0.19 for P4. The density loss expected under this expansion, as many added entities are described by only one source, is nearly identical at $-0.09$, $-0.09$, and $-0.08$. The larger output therefore does not come at the cost of a much sparser table. The main structural difference is clustering: P4 grows the record count least but fuses most often, with a mean fusion ratio of 0.46 against 0.39 for P2 and 0.35 for P3. The per-task spread of this ratio reflects task structure and matcher behavior. On Products, where the four sources describe largely overlapping offers, 0.84 (P2), 0.69 (P3), and 0.94 (P4) of the output records combine several sources. On Games, where most entities appear in only one source, the ratio stays at 0.11, 0.17, and 0.31. On a high-overlap task, a low ratio instead signals a matcher that leaves records as singletons. P3 has the lowest fusion ratio on Products, and the ground-truth rows below confirm that it splits these entities, with a BCubed recall of 0.44. The reference-free view flags structural differences but cannot judge correctness.

Mean schema validity is high for all three pipelines, at 0.95 for P2, 0.98 for P3, and 1.00 for P4. High results such as the 1.00 of all three pipelines on Games are possible as strict checks cover dates, scores, ESRB labels, and list length, while the genre and platform taxonomies are intentionally non-exhaustive. A non-exhaustive taxonomy does not contain every possible value, so the validity check does not penalize values outside the taxonomy. P2 reaches lower validity on Music (0.84) and Companies (0.94). On Companies, most invalid cells are industry and country values outside the target taxonomies (2,874 of 3,114 invalid cells). On Music, genre values outside the genre taxonomy (14,193 cells), release dates that give only a year or a year and month instead of a full date (13,286), and empty track lists or track names (5,504) account for almost all invalid cells.
The rows below quantify how much this appearance hides: the same outputs that reach 0.95 to 1.00 schema validity reach ground-truth fully-correct rates of 0.10 to 0.25. Reference-free metrics confirm target-format compatibility, and the reference-based rows are needed to evaluate factual quality.

\textbf{Silver reference:} P4 is closest to the P1 clustering, with 0.61 entity recovery and 0.82 BCubed F1, against 0.56 and 0.80 for P2 and 0.47 and 0.74 for P3. The difference to P2 comes mainly from Games (entity recovery 0.78 against 0.50). P2 is closest on fused values, with a fusion accuracy of 0.71 against 0.69 for P4 and 0.62 for P3, and a fully-correct rate of 0.22 against 0.14 and 0.05, mainly due to Papers (0.55 against 0.13 and 0.01). P4 thus more often recreates the P1 grouping, while P2 more often selects the same values once an entity is aligned.

\textbf{Ground-truth reference:} On the test sets, P4 yields the best clustering result. It recovers 0.87 of the labeled entities on average, compared with 0.77 for P2 and 0.58 for P3. The gap is largest on Products (0.76, 0.59, and 0.08), where BCubed F1 follows at 0.94, 0.89, and 0.59. The value-level metrics show a different view of the results. P3 has a lower value drift than P2 (0.43 against 0.46) but also a lower fusion accuracy (0.72 against 0.77). Matching the aggregate value distribution therefore does not imply assigning the correct values to the correct entities. P4 reaches both the lowest drift (0.40) and the highest fusion accuracy (0.83).

The ground-truth reference also reverses the silver ranking of P2 and P4 on fusion accuracy. On the silver reference, P2 reaches the higher fusion accuracy (0.71 against 0.69). On the ground truth, P4 does (0.83 against 0.77), and it also reaches the higher fully-correct rate (0.25 against 0.17). Agreement with a reference pipeline and correctness on independently verified records thus capture two different sides of the fused result. Possibly, a silver reference favors systems resembling the reference pipeline (Section~\ref{sec:eval-discussion}).

The strict fully-correct rate remains low for all pipelines, at 0.17 for P2, 0.10 for P3, and 0.25 for P4 on average, and at most 0.36 on any task (P2 on Papers, P4 on Music). This metric requires every evaluated attribute of an entity to be correct at once, so it compounds the per-step errors. Even P4, with a mean ground-truth entity recovery of 0.87, produces a fully correct record for only about one in four test entities, which shows the difficulty of fusion.

\textbf{Runtime:} Table~\ref{tab:runtime-results} reports wall-clock runtime in seconds for the four pipelines. All values measure the execution of the final pipeline. Model training is part of the measured runtime, while the configuration search and the human design effort are not.

Mean runtimes differ by a factor of 15, from 358 seconds for P4 to 5{,}205 seconds for P2, with 469 for P3 and 834 for P1. The two steps with the largest influence on runtime are blocking and entity matching. Papers, the largest task, is the most expensive task for every pipeline, even at reduction ratios of at least 99.97 (Table~\ref{tab:blocking-em-results}).
Most of the P2 runtime is the per-task fine-tuning and inference of Ditto. On Companies, the only task on which the committee selects Magellan, P2 needs 455 seconds, its shortest runtime. On Papers, P2 spends about 7.5 times the runtime of P1 for 1.39 additional matching F1 (Table~\ref{tab:blocking-em-results}).

\textbf{Cost:} Table~\ref{tab:runtime-results} also reports the estimated API cost of the two LLM-driven pipelines. P3 and P4 cost about four US dollars per task on average (3.96 and 3.86), with ranges of 0.87 to 6.70 for P3 and 2.47 to 5.95 for P4. The two pipelines spend their budgets differently. For P3, labeling and training data creation is the most expensive step, with 64.8\% of the recorded LLM cost, followed by the selection and optimization of the fusion rules with 16.4\%. P4 spends its budget while the agent writes, tests, and revises the pipeline code. A post-hoc assignment of the agent's responses to integration steps attributes about 45\% of the P4 cost to blocking, entity matching, and clustering. P3 directs most of its budget to training labels, while P4 directs it to developing the pipeline itself. 

\begin{table}[t]
\centering
\caption{Runtime (s) and estimated API cost (USD).}
\label{tab:runtime-results}
\scriptsize
\setlength{\tabcolsep}{4pt}
\begin{tabular}{@{}lrrrrrr@{}}
\toprule
Task & \shortstack{Human\\(P1)} & \shortstack{BoB\\(P2)} & \shortstack{LLM\\(P3)} & \shortstack{CC\\(P4)} & \shortstack{LLM (P3)\\cost (USD)} & \shortstack{CC (P4)\\cost (USD)} \\
\midrule
Games & 789 & 2612 & 196 & 374 & 3.17 & 3.06 \\
Companies & 269 & 455 & 203 & 180 & 0.87 & 4.22 \\
Music & 728 & 7231 & 211 & 290 & 3.25 & 3.58 \\
Products & 441 & 1029 & 363 & 22 & 5.79 & 5.95 \\
Papers & 1944 & 14699 & 1374 & 922 & 6.70 & 2.47 \\
\midrule
Mean & 834 & 5205 & 469 & 358 & 3.96 & 3.86 \\
\bottomrule
\end{tabular}
\end{table}

\subsection{Discussion}\label{sec:eval-discussion}

The validation shows four properties of the benchmark. First, no pipeline performs best on every step and task. Entity matching, normalization, and fusion each rank the pipelines differently per task. Second, the base tasks leave room for improvement. Every pipeline scores below 40 normalization accuracy on at least one task, and the mean ground-truth fully-correct rate stays at or below 0.25. Claude Code with Claude Opus 5.5 (P4) reaches the highest fusion accuracy on four of five tasks but leaves 13.36 to 22.19 points open on every task. Third, the end-to-end metrics add information beyond the step-wise scores. The silver and ground-truth references rank P2 and P4 differently on fusion accuracy, and high schema validity coexists with low fully-correct rates. Fourth, the mean runtimes differ by a factor of 15, so the effectiveness of a pipeline can be weighed against its runtime.

Two limitations apply. First, the silver reference is produced by the human PyDI pipeline, so silver metrics may favor systems whose decisions resemble PyDI-style workflows. The human-verified test sets are independent of a specific pipeline and give a stronger correctness estimate, although they only cover a small subset of the overall output dataset. Second, the source data is partly public, so LLM-based systems may have seen parts of it during pretraining. The perturbed task variants introduced in the next section reduce the value of such memorization.

\section{Variant Generation}
\label{sec:taskgen}

The five base tasks are real-world datasets having a fixed level of difficulty. Given the current pace of improvement in the area of agentic end-to-end data integration systems, it is likely that the benchmark tasks get saturated quickly, making them unsuitable for evaluation of data integration systems in the long-run. On the other hand, the current difficulty of some of the tasks may already be higher than what a practitioner may face on tasks in their domain, making the application of simpler less costly methods relevant, which may be unsuited for the benchmarks base tasks, while they may prove to be effective and efficient alternatives for the use cases faced by the practitioners. In order to address both of these cases, we derive three \emph{variants} of each base task, namely \emph{easy}, \emph{medium}, and \emph{hard}. The variants hold the underlying integration problem fixed but vary the level of difficulty of the subtasks. Because the variants amplify or reduce the same forms of heterogeneity that make real integration hard, a harder or easier level changes the difficulty of the same integration problem. As a result, the benchmark stays future-proof as systems improve, while also supporting less demanding scenarios where efficiency matters most. %

The variants differ from the base tasks in how hard the same integration problem is to solve. Across the \emph{easy}, \emph{medium}, and \emph{hard} levels, we reduce surface overlap by injecting value noise such as typos and encoding artefacts, add negative and positive corner-case records to the clusters describing specific entities, drop cell values and skew which sources cover each entity, diversify date, number, and unit formats, rename source columns to synonyms or opaque codes, and shuffle which source holds the correct fusion value so that no single source is more trustworthy in general. Importantly, these changes never alter the correspondences among the underlying records, a match on easy does not become a non-match on hard due to value perturbation, only the difficulty is adjusted so the decision moves further or closer to the boundary between matches and non-matches. As a result, a single dataset spans a controlled difficulty range that affects all steps of the integration workflow.

The augmentation method that derives the variants perturbs the source data along a set of independently controllable \emph{difficulty knobs}, each set to an absolute target for the three difficulty levels, for example, a corner-case ratio of 20/50/80\% for easy/medium/hard or a per attribute drop rate of 5/15/35\%, and each targeting a specific challenge of the integration task. As a result, a variant can add or remove heterogeneity relative to the domain's base task (Section~\ref{sec:taskgen:knobs}). The knobs are applied in a fixed order so that the perturbations compose predictably, while some entities and values are protected to ensure the evaluation sets remain useable (Section~\ref{sec:taskgen:method}). After creation, a validation step measures whether the resulting variants are harder or easier by running heterogeneous algorithm committees for each pipeline stage and reporting the committee mean and ceiling across difficulty levels and against the base task (Section~\ref{sec:taskgen:validation}). Section~\ref{sec:taskgen:validation} also reports the variant statistics, and Section~\ref{sec:taskgen:results} reports best-of-breed, LLM-based, and Claude Code pipeline results on the variants relative to the base tasks.

\subsection{Difficulty Knobs}
\label{sec:taskgen:knobs}
\begin{table}[t]
\centering
\caption{A DBpedia Companies record at base and \emph{hard} difficulty. Colors mark the responsible knob: \textcolor{teal}{schema naming divergence (7)}, \textcolor{blue}{surface augmentation (1)}, \textcolor{orange}{format and unit diversity (5)}, \textcolor{purple}{value noise (6)}, and \textcolor{gray}{attribute drop (3)}. }
\label{tab:knob-example}
\scriptsize
\setlength{\tabcolsep}{3pt}
\begin{tabular}{@{}lll@{}}
\toprule
Attribute (base $\rightarrow$ hard) & Base value & Hard value \\
\midrule
\texttt{org\_name} $\rightarrow$ \textcolor{teal}{\texttt{nm}} & Taisei Corporation & \textcolor{blue}{Taisei Corp} \\
\texttt{established} $\rightarrow$ \textcolor{teal}{\texttt{ey}} & 1873-01-01 & \textcolor{orange}{01.01.187312:00:00} \\
\texttt{nation} $\rightarrow$ \textcolor{teal}{\texttt{cn}} & Japan & \textcolor{blue}{Nippon} \\
\texttt{sector} $\rightarrow$ \textcolor{teal}{\texttt{sg}} & Construction & \textcolor{purple}{Constructi on} \\
\texttt{keypeople\_name} $\rightarrow$ \textcolor{teal}{\texttt{kpn}} & Okura Kihachiro & \textcolor{gray}{\emph{dropped}} \\
\texttt{total\_assets\_val} $\rightarrow$ \textcolor{teal}{\texttt{ta}} & 3650187000 & \textcolor{orange}{3650.19} \\
\texttt{annual\_income} $\rightarrow$ \textcolor{teal}{\texttt{ai}} & 213195000 & \textcolor{orange}{167.18} \\
\bottomrule
\end{tabular}
\end{table}

We control the difficulty of an end-to-end DI task using eight knobs. Each knob is an independently togglable perturbation that targets one or more difficulty dimensions of a specific pipeline step, with absolute targets per \emph{easy}/\emph{medium}/\emph{hard} level. We describe the eight knobs in turn, naming in parentheses the difficulty dimension that each effect targets. Table~\ref{tab:knob-example} illustrates the value-level knobs on a single Companies record at \emph{hard} difficulty.

\textbf{1) Surface augmentation intensity} rewrites values into alternative surface forms using abbreviations, token reorderings, reformulations, and dropped tokens, eroding the surface overlap that blocking and matching rely on (representation heterogeneity, corner-case difficulty) and yielding plausible value disagreements for fusion (conflict subtlety). The per-value rewrite probability is 0/4/12\% for names and titles, 0/8/18\% for key and categorical attributes, and 0/16/30\% for the other attributes.

\textbf{2) Entity niche density} adds similar-but-distinct entities within the same domain niche, such as many products of one brand, enlarging blocks with non-matching pairs (candidate density) and pushing candidate pairs toward the matching decision boundary (corner-case ratio). The corner-case ratio is set to 20/50/80\%.

\textbf{3) Per-source attribute drop rate} blanks 5/15/35\% of cells per source, removing blocking keys (blocking-key completeness), matching evidence (record completeness), and the redundancy that fusion cross-validates against (source density).

\textbf{4) Per-entity coverage skew} varies how many sources describe each entity. The share of single-source entities rises over 0/20/55\%, skewing group sizes and thinning the evidence for fusion (source density, conflict rate).

\textbf{5) Format and unit diversity} rewrites dates, numbers, and quantities into two to four admissible formats per type and one to three locales, mixing units and scales at harder levels (format heterogeneity, unit and scale diversity).

\textbf{6) Value-noise injection rate} corrupts 1/4/12\% of values with typos, encoding artefacts, and truncation from the FEBRL corruption taxonomy~\cite{christenFebrlOpenSource2008}, degrading similarity signals for matching and surfacing as source disagreements for fusion (noise and corruption, corner-case difficulty, conflict rate).

\textbf{7) Schema naming divergence} renames source columns from fully descriptive (easy) through abbreviations (medium) to opaque codes (hard), removing the attribute name signal that schema matchers rely on (naming heterogeneity).

\textbf{8) Source reliability differentiation} reshuffles which source holds the correct value per attribute. The most reliable source holds it for 85/65/50\% of an attribute's entities, so that no fixed source-trust strategy stays optimal (trust ambiguity, conflict rate).

\subsection{Generation Method}
\label{sec:taskgen:method}

The following paragraphs describe the relevant artefacts used in the creation of the variants.

\textbf{Positive Pool:} The variant generator must be able to perturb the data without destroying the existing cross-source matches, i.e. it should not change a match into a non-match. The labeled entity matching training, validation and test sets cover only a small sample of all true pairs. Therefore, we build a per-domain \emph{positive pool}: an approximation of the real matches, assembled by merging the labeled positives, the correspondences emitted by each domain's human-built baseline matcher, and the predictions of a domain-trained Ditto~\cite{ditto} matcher over a blocker-derived candidate set. Pairs on which the baseline matcher and Ditto agree enter the pool directly. The remaining disagreements are arbitrated by an LLM (GPT-5.4), and the pool is closed transitively across source pairs so that a match inferred for one pair propagates to the others. For \emph{products} and \emph{papers}, the existing \emph{cluster\_id} and \emph{DOI} attributes provide full linkage, so we build the positive pool directly.

\textbf{Protected Entities and Values:} The entities in the fusion validation and test sets form a protection set. No knob may remove all records of such an entity, and Entity niche density (Knob~2) may not touch any of their records. Thus, every entity referenced by the fusion validation or test set survives at every difficulty level.

A strong perturbation could still corrupt every source value of a fused cell beyond recovery. To handle this, for each fusion value, the generator enforces a \emph{closeness floor}: at least one surviving source value for an entity must stay within an attribute-specific tolerance of the target value ($\pm 3\%$ for continuous quantities, $\pm 1$ year for years and $\pm 1$ day for full dates, and normalized Levenshtein or token-Jaccard similarity thresholds of $0.5$-$0.85$ for strings), so that a suitable fusion strategy can still recover the correct value. The value-corruption knobs (surface augmentation and value noise) mutations are discarded whenever they would push the last source value of an entity out of tolerance. Format and unit changes (Knob~5) are exempt, as they are equivalence transforms that can be reverted to the same fused value. The target values used for protection come from the fusion validation and test sets. To extend protection beyond those two hundred verified entities for each domain to further entities, we also use a \emph{fusion silver standard}. This silver standard applies the fusion rules of each domain's hand-built pipeline (P1) to the clusters of the positive pool. It extends the protection to every entity that the pool links across sources.

\textbf{Evaluation Sets:} The fusion validation and test sets keep their entities and gold values at every level, while the knobs perturb the source records of these entities. Where knobs removed some of an entity's records (mostly Knob~4 on \emph{hard}), the gold record lists only the surviving records and, if its key record was removed, is keyed to a surviving one. The entity matching evaluation sets are copied from the base tasks. Any record pairs that no longer exist due to dropped records are removed and subsequently filled with positive and negative corner-cases until they have a similar size and positive ratio as the base task (Section~\ref{sec:taskgen:validation}). The schema matching test set is rewritten to fit the Knob~7 renames. Each variant also provides normalization validation and test sets on its own source values.

\subsection{Variant Validation and Statistics}
\label{sec:taskgen:validation}

To assess the difficulty of the variants, we measure the changes on each step of the integration pipeline in comparison to the base tasks. For each stage of the pipeline, we run a \emph{committee} of algorithms (Table~\ref{tab:committees}) on every difficulty level. Every member is fitted once on the base task and then scored unchanged on each variant.

\begin{table}[t]
\centering
\caption{Committee members per pipeline stage.}
\label{tab:committees}
\footnotesize
\setlength{\tabcolsep}{5pt}
\begin{tabular}{@{}l p{5.7cm}@{}}
\toprule
Stage & Committee members \\
\midrule
Schema matching (7) & label-based, instance-based and duplicate-based matching, Sentence-BERT~\cite{sbert}, COMA~\cite{coma}, Magneto~\cite{magneto}, GPT-5.4-mini \\
Normalization (3) & rule-based per-attribute selection (text cleaning, date and locale-aware number parsing, country codes, taxonomy lookup), LLM canonicalization (GPT-5.4-mini), passthrough \\
Blocking (6) & standard, token, sorted-neighborhood~\cite{christophides2020matchig}, Sentence-BERT~\cite{sbert}, BM25~\cite{bm25}, SC-Block~\cite{scblock} \\
Entity matching (4) & Magellan~\cite{magellan}, Ditto~\cite{ditto}, MatchGPT~\cite{peeters2025entitymatchingusinglarge}, ComEM~\cite{wangMatchCompareSelect2025} \\
Data fusion (8) & Voting, source-trust-based, PyDI per-attribute, TruthFinder~\cite{yinTruthDiscoveryMultiple2007}, AccuSim~\cite{dongIntegratingConflictingData2009}, LTM~\cite{zhaoBayesianApproachDiscovering2012}, FusionQuery~\cite{zhuFusionQueryOndemandFusion2024}, CASEFusion~\cite{lyuTruthDiscoveryClaim2017} \\
\bottomrule
\end{tabular}
\end{table}

Each committee reports two measurements: the \emph{mean}, the macro-average over all members, and the \emph{ceiling}, the score of the strongest member at each level of difficulty. Raising the difficulty should depress both, so that a variant challenges not only the average member but also the best method. Table~\ref{tab:variant-stats} reports both metrics per stage, alongside the variant statistics. On easy and medium levels, row counts stay within 7\% of the base task. The hard variants of Products and Papers are about a fifth and a quarter smaller. Source density is lower on every variant than on the base task.

The easy committee mean exceeds the base score for schema matching in all five domains and for entity matching in four of five, so the \emph{easy} level lowers the difficulty of these steps. Fusion is the exception, with an easy mean below the base score in four domains, as attribute drop and value noise already apply at this level. From \emph{medium} to \emph{hard}, the committee mean declines for schema matching, entity matching, and fusion in every domain. The hard mean lies below the base score for entity matching and fusion in every domain and for schema matching in four of five. The signal is clearest for fusion, whose mean falls by 10.9 to 24.0 points and whose ceiling by 10.6 to 28.1 points from base to hard.

For schema matching the mean and the ceiling diverge: the mean drops under the header renames of Knob~7, which collapse the label-focused matchers, while the strongest instance- and LLM-based matchers still recover the mapping from the values. The entity matching mean declines from easy to hard in every domain by 3.1 to 9.1 points, while the ceiling changes by at most 5.3 points.

\begin{table}[!t]
\centering
\caption{Variant statistics and committee difficulty signal.}
\label{tab:variant-stats}
\scriptsize
\setlength{\tabcolsep}{3pt}
\resizebox{\columnwidth}{!}{%
\begin{tabular}{@{}ll cc cccc cccc@{}}
\toprule
& & \multicolumn{2}{c}{Sources} & \multicolumn{4}{c}{\shortstack{Committee\\mean (\%)}} & \multicolumn{4}{c}{\shortstack{Best-member\\ceiling (\%)}} \\
\cmidrule(lr){3-4}\cmidrule(lr){5-8}\cmidrule(lr){9-12}
Domain & Lvl & Rows\,(k) & Density & SM & Norm. & EM & Fus. & SM & Norm. & EM & Fus. \\
\midrule
\multirow{4}{*}{Companies} & Base & 14.0 & 66 & 68.6 & 35.9 & 88.4 & 56.9 & 100.0 & 43.5 & 89.5 & 62.4 \\
 & Easy & 14.0 & 63 & 88.5 & 20.3 & 89.9 & 41.1 & 100.0 & 22.2 & 93.4 & 44.6 \\
 & Med. & 14.3 & 59 & 79.5 & 27.2 & 87.2 & 49.3 & 100.0 & 31.2 & 88.2 & 53.3 \\
 & Hard & 14.2 & 58 & 66.4 & 29.9 & 86.8 & 37.5 & 100.0 & 34.6 & 88.1 & 38.6 \\
\midrule
\multirow{4}{*}{Games} & Base & 75.0 & 94 & 69.6 & 8.1 & 60.9 & 68.1 & 100.0 & 12.1 & 71.6 & 70.7 \\
 & Easy & 78.5 & 84 & 88.1 & 7.0 & 63.3 & 66.8 & 100.0 & 9.8 & 71.8 & 68.9 \\
 & Med. & 75.3 & 83 & 78.5 & 19.9 & 64.8 & 63.4 & 100.0 & 24.4 & 75.5 & 66.2 \\
 & Hard & 71.1 & 83 & 69.3 & 22.4 & 55.3 & 57.2 & 100.0 & 25.0 & 71.5 & 60.1 \\
\midrule
\multirow{4}{*}{Music} & Base & 37.3 & 90 & 67.0 & 48.8 & 82.4 & 74.5 & 100.0 & 61.3 & 95.3 & 82.0 \\
 & Easy & 39.7 & 86 & 91.8 & 23.2 & 88.3 & 75.5 & 100.0 & 29.3 & 95.1 & 80.4 \\
 & Med. & 37.3 & 82 & 85.4 & 46.8 & 82.7 & 55.8 & 100.0 & 61.6 & 94.8 & 62.0 \\
 & Hard & 35.2 & 80 & 77.3 & 40.5 & 79.2 & 50.5 & 100.0 & 53.1 & 92.9 & 53.9 \\
\midrule
\multirow{4}{*}{Products} & Base & 3.0 & 53 & 71.7 & 45.7 & 89.7 & 80.4 & 100.0 & 85.2 & 95.5 & 82.8 \\
 & Easy & 3.1 & 47 & 79.1 & 45.1 & 92.6 & 78.6 & 100.0 & 86.6 & 96.4 & 81.8 \\
 & Med. & 3.1 & 42 & 72.1 & 47.0 & 88.9 & 71.2 & 100.0 & 90.7 & 96.6 & 76.6 \\
 & Hard & 2.4 & 45 & 65.7 & 46.6 & 85.8 & 60.9 & 98.0 & 91.3 & 97.4 & 65.3 \\
\midrule
\multirow{4}{*}{Papers} & Base & 182.1 & 82 & 75.0 & 35.6 & 96.6 & 81.9 & 98.5 & 66.9 & 100.0 & 92.3 \\
 & Easy & 182.2 & 64 & 82.3 & 35.0 & 96.5 & 79.4 & 100.0 & 64.8 & 99.9 & 89.5 \\
 & Med. & 182.1 & 60 & 74.9 & 35.7 & 95.3 & 77.3 & 100.0 & 68.0 & 99.9 & 86.1 \\
 & Hard & 136.9 & 57 & 62.9 & 36.8 & 92.2 & 66.6 & 98.6 & 70.3 & 99.4 & 70.6 \\
\bottomrule
\end{tabular}}
\end{table}

\subsection{Variant Evaluation}
\label{sec:taskgen:results}

Table~\ref{tab:variant-results} reports the per-stage scores of P2, P3, and P4 on each base task and its three variants. Fusion shows the clearest difficulty gradient. For every pipeline and domain, fusion accuracy on the hard variant lies below the base task, with losses of 16.7 to 30.0 points for P2, 13.1 to 29.7 for P3, and 2.0 to 14.6 for P4. The hard variant is the level with the lowest fusion accuracy in 14 of the 15 combinations of pipeline and domain.

P4 reaches the highest base fusion accuracy on Companies, Games, Music, and Products. The hard variants lower it most on Music, from 86.6 to 72.0, and on Papers, from 85.6 to 71.6. On the hard variants, no pipeline exceeds 75.8 fusion accuracy in any domain. The easy variants do not raise fusion accuracy in general. On Companies, P2 and P3 lose 17.2 and 23.1 points from base to easy, which matches the lower easy fusion mean of the committees.

Schema matching stays at or above 91.3 F1 on every level. Entity matching separates the pipelines more. On Products, P3 drops to 20.9 F1 on medium, while P2 and P4 stay above 79 F1 throughout.

\begin{table}[!t]
\centering
\caption{Per-stage scores on the base task and its variants.}
\label{tab:variant-results}
\scriptsize
\setlength{\tabcolsep}{3pt}
\resizebox{\columnwidth}{!}{%
\begin{tabular}{@{}ll rrrr rrrr rrrr@{}}
\toprule
& & \multicolumn{4}{c}{BoB (P2)} & \multicolumn{4}{c}{LLM (P3)} & \multicolumn{4}{c}{CC (P4)} \\
\cmidrule(lr){3-6}\cmidrule(lr){7-10}\cmidrule(lr){11-14}
Domain & Lvl & SM & Norm. & EM & Fus. & SM & Norm. & EM & Fus. & SM & Norm. & EM & Fus. \\
\midrule
\multirow{4}{*}{Companies} & Base & 97.7 & 29.3 & 89.3 & 66.4 & 92.7 & 84.8 & 92.6 & 75.7 & 100.0 & 97.8 & 91.2 & 82.0 \\
 & Easy & 100.0 & 22.2 & 93.6 & 49.2 & 100.0 & 57.6 & 90.0 & 52.6 & 100.0 & 73.7 & 96.0 & 81.5 \\
 & Med. & 100.0 & 29.9 & 88.9 & 58.5 & 100.0 & 65.0 & 77.4 & 62.8 & 100.0 & 86.6 & 90.2 & 79.8 \\
 & Hard & 91.3 & 24.4 & 94.5 & 40.6 & 100.0 & 64.7 & 88.5 & 51.1 & 100.0 & 59.6 & 95.1 & 71.2 \\
\midrule
\multirow{4}{*}{Games} & Base & 100.0 & 6.0 & 67.3 & 65.7 & 100.0 & 37.9 & 65.0 & 71.4 & 93.9 & 8.8 & 85.4 & 77.8 \\
 & Easy & 100.0 & 9.8 & 66.9 & 62.7 & 100.0 & 43.7 & 78.8 & 72.8 & 100.0 & 44.7 & 94.1 & 81.1 \\
 & Med. & 100.0 & 24.4 & 69.7 & 62.4 & 100.0 & 50.0 & 72.2 & 70.0 & 100.0 & 36.7 & 79.3 & 71.6 \\
 & Hard & 100.0 & 22.4 & 79.2 & 49.0 & 100.0 & 50.4 & 81.3 & 58.3 & 100.0 & 51.8 & 92.7 & 75.8 \\
\midrule
\multirow{4}{*}{Music} & Base & 100.0 & 55.0 & 94.8 & 81.4 & 100.0 & 75.0 & 98.1 & 75.2 & 100.0 & 56.3 & 95.5 & 86.6 \\
 & Easy & 100.0 & 17.4 & 94.9 & 75.2 & 100.0 & 20.7 & 97.8 & 69.1 & 100.0 & 21.7 & 95.3 & 79.8 \\
 & Med. & 100.0 & 56.5 & 93.2 & 61.0 & 100.0 & 19.4 & 90.7 & 64.9 & 100.0 & 67.1 & 96.2 & 81.4 \\
 & Hard & 100.0 & 42.0 & 94.3 & 53.9 & 100.0 & 19.1 & 88.2 & 53.2 & 100.0 & 64.8 & 96.5 & 72.0 \\
\midrule
\multirow{4}{*}{Products} & Base & 100.0 & 22.0 & 85.5 & 79.3 & 100.0 & 30.0 & 66.0 & 61.5 & 100.0 & 73.2 & 88.5 & 82.8 \\
 & Easy & 100.0 & 18.8 & 96.9 & 81.6 & 100.0 & 29.8 & 51.2 & 69.1 & 100.0 & 53.1 & 95.9 & 85.9 \\
 & Med. & 100.0 & 20.4 & 86.0 & 73.9 & 100.0 & 29.9 & 20.9 & 54.9 & 100.0 & 69.0 & 85.0 & 75.2 \\
 & Hard & 98.0 & 18.5 & 79.2 & 54.4 & 100.0 & 30.0 & 50.1 & 37.0 & 100.0 & 53.1 & 91.7 & 74.7 \\
\midrule
\multirow{4}{*}{Papers} & Base & 100.0 & 29.8 & 99.7 & 90.2 & 100.0 & 29.8 & 96.8 & 78.4 & 100.0 & 64.5 & 99.8 & 85.6 \\
 & Easy & 100.0 & 40.2 & 99.5 & 88.0 & 100.0 & 40.2 & 90.6 & 72.1 & 100.0 & 72.3 & 99.6 & 84.8 \\
 & Med. & 98.6 & 39.0 & 99.2 & 85.2 & 100.0 & 39.0 & 87.2 & 69.8 & 100.0 & 75.7 & 99.6 & 81.8 \\
 & Hard & 100.0 & 39.1 & 96.7 & 60.2 & 100.0 & 43.2 & 77.7 & 48.7 & 100.0 & 51.0 & 95.1 & 71.6 \\
\bottomrule
\end{tabular}}
\end{table}

In summary, the \emph{hard} variants lower the fusion accuracy of all three pipelines in every domain and leave room for future systems to improve. If future systems saturate the hard variants, the knob targets (Section~\ref{sec:taskgen:knobs}) can be raised to derive harder variants. The \emph{easy} variants lower the difficulty of schema and entity matching and target simpler and computationally cheaper methods.

\section{Related Work}\label{sec:related}

\begin{table}[t]
\centering
\caption{Evaluation scope of related benchmarks: efficiency (Eff.), end-to-end output quality (E2E), and step-level evaluation of schema matching (SM), blocking (Bl.), entity matching (EM), and data fusion (Fus.). Normalization is not compared. \(\checkmark\): yes, \(\circ\): partial, \(\times\): no.}
\label{tab:related-benchmark-comparison}
\scriptsize
\setlength{\tabcolsep}{2.5pt}
\renewcommand{\arraystretch}{1.08}
\resizebox{\columnwidth}{!}{%
\begin{tabular}{@{}llcccccc@{}}
\toprule
\textbf{Benchmark} & \textbf{Eval. unit} & \textbf{Eff.} & \textbf{E2E} & \textbf{SM} & \textbf{Bl.} & \textbf{EM} & \textbf{Fus.} \\
\midrule
TPC-DI~\cite{tpc-di} & warehouse load & \(\checkmark\) & \(\times\) & \(\times\) & \(\times\) & \(\times\) & \(\times\) \\
DIPBench~\cite{dipbench} & process execution & \(\checkmark\) & \(\times\) & \(\times\) & \(\times\) & \(\times\) & \(\times\) \\
DIBS~\cite{dibs} & preprocessing kernels & \(\checkmark\) & \(\times\) & \(\times\) & \(\times\) & \(\times\) & \(\times\) \\
Alaska~\cite{alaska} & schema matching/ER & \(\times\) & \(\times\) & \(\checkmark\) & \(\times\) & \(\checkmark\) & \(\times\) \\
LakeBench~\cite{lakebench} & join/union discovery & \(\checkmark\) & \(\times\) & \(\times\) & \(\times\) & \(\times\) & \(\times\) \\
ELT-Bench~\cite{elt-bench} & agent ELT pipeline & \(\checkmark\) & \(\checkmark\) & \(\times\) & \(\times\) & \(\times\) & \(\times\) \\
KGI-Bench~\cite{kgi-bench, hofer2150kgpipe} & knowledge graph & \(\checkmark\) & \(\checkmark\) & \(\checkmark\) & \(\times\) & \(\checkmark\) & \(\circ\) \\
MaDI-Bench & fused table & \(\checkmark\) & \(\checkmark\) & \(\checkmark\) & \(\checkmark\) & \(\checkmark\) & \(\checkmark\) \\
\bottomrule
\end{tabular}
}
\end{table}

\balance

This section compares MaDI-Bench to existing benchmarks in data integration and data engineering.

TPC-DI~\cite{tpc-di}, DIPBench~\cite{dipbench}, and DIBS~\cite{dibs} measure the efficiency of integration systems. TPC-DI models a retail-brokerage ETL workload with fixed source and target models, transformations, run rules, and a throughput metric, so it evaluates how efficiently a system executes the specified load rather than whether it can infer integration decisions. DIPBench defines source and target schemas, 15 integration process types, concurrent execution streams, and scale factors for measuring integration-process execution, while DIBS provides preprocessing kernels for parsing, transformation, and aggregation and characterizes their hardware behavior.

Alaska~\cite{alaska} exposes schema matching and entity resolution tasks over about 70k Web product specifications with manually curated ground truth. LakeBench~\cite{lakebench} evaluates the effectiveness and efficiency of top-\(k\) discovery of joinable and unionable tables over more than 16 million tables and more than 10{,}000 labeled queries, addressing discovery before integration.

ELT-Bench~\cite{elt-bench} and KGI-Bench evaluate complete pipelines but target different artefacts. ELT-Bench evaluates whether agents can configure Airbyte/Snowflake pipelines and write dbt transformations for target data models, reporting stage success, target-model correctness, cost, and steps. KGI-Bench~\cite{kgi-bench,hofer2150kgpipe} evaluates pipelines for integrating data into knowledge graphs. Similar to MaDI-Bench, KGI-Bench covers schema alignment, entity resolution, and entity fusion, but it covers a single domain: integrating movie data, given as RDF, JSON, and text, into a seed knowledge graph derived from DBpedia. A main difference between the benchmarks is that MaDI-Bench addresses the integration of relational tables, while KGI-Bench addresses the augmentation of RDF knowledge graphs.

Table~\ref{tab:related-benchmark-comparison} compares the different benchmarks. The efficiency-oriented benchmarks score none of the integration decisions. Step-level evaluation is limited to Alaska's schema matching and entity resolution ground truth and KGI-Bench for the task of augmenting knowledge graphs. The comparison shows that MaDI-Bench is the only benchmark covering all steps of the integration process for the use case of integrating tabular data.

\section{Conclusion}\label{sec:conclusion}

This paper introduced the Mannheim Data Integration Benchmark (MaDI-Bench). The benchmark covers all steps of the data integration process, from schema matching and value normalization through blocking and entity matching to data fusion, and evaluates each step as well as the integrated output, taking into account the dependencies between the steps.
To prevent a quick saturation of the benchmark as agentic systems progress, we introduced a generic variant-generation method for deriving harder as well as easier variants from the base tasks. We validated MaDI-Bench and the generated variants using human-designed pipelines, an LLM workflow, a best-of-breed pipeline, and a Claude Code integration agent. The validation showed the benchmark's utility for measuring the step-wise as well as end-to-end performance of data integration pipelines. We hope that the benchmark will prove useful for the community and will support the development of fully automatic as well as human in the loop data integration systems.

\begin{acks}
The authors used Claude Code and OpenAI Codex to develop code and revise the text, and OpenAI's image-generation tool to draw Figure~\ref{fig:workflow}. The authors reviewed all AI-assisted content and take full responsibility for it. The design of the benchmark, its motivation, and the interpretation of the experimental results are exclusively the authors' own.
\end{acks}

\bibliographystyle{ACM-Reference-Format}
\bibliography{references}

\end{document}